\documentclass{article}
\usepackage{amssymb}
\usepackage{graphics}
\usepackage{opex2}

\begin{document}

\newcommand{\ket}[1]{|{#1}\rangle}
\newcommand{\bra}[1]{\langle{#1}|}
\newcommand{\braket}[1]{\langle{#1}\rangle}
\newcommand{\ad}{a^\dagger}

\title{Variational ansatz for the superfluid Mott-insulator transition in optical lattices}
\author{J.~J. Garc\'{\i}a-Ripoll and J.~I. Cirac}
\address{Max-Planck-Institut f\"ur Quantenoptik,
  Hans-Kopfermann-Str. 1, Garching, D-85748, Germany.}
\email{Juan.Ripoll@mpq.mpg.de}
\author{P. Zoller}
\address{Leopold-Franzens Universit\"at Innsbruck, Technikerstr. 24, Innsbruck, A-6020, Austria.}
\author{C. Kollath, U. Schollw\"ock and J. von Delft}
\address{Ludwig-Maximilians-Universit\"at, Theresienstr. 37, D-80333
  M\"unchen, Germany.}

\begin{abstract}
  We develop a variational wave function for the ground state of a
  one-dimensional bosonic lattice gas. The variational theory is initally
  developed for the quantum rotor model and later on extended to the
  Bose-Hubbard model. This theory is compared with quasi-exact numerical
  results obtained by Density Matrix Renormalization Group (DMRG) studies and
  with results from other analytical approximations.  Our approach accurately
  gives local properties for strong and weak interactions,
  and it also describes the crossover from
  the superfluid phase to the Mott-insulator phase.
\end{abstract}

%% 03.75.Kk Matter waves
%% 05.30.Jp Boson systems
%% 03.75 Lm Tunneling, Josephson effect, Bose-Einstein condensates in
%%          periodic potentials, solitons, vortices and topological excitations
%% 73.43.Nq Quantum phase transitions
%%\pacs{03.75.Kk,73.43.Nq,05.30.Jp,03.75.Lm}

%\bibliography{references}

%\end{OEReferences}

% ============================================================

\section{ Introduction}

During the last years a spectacular development in the storage and
manipulation of cold atoms in optical lattices \cite{Bloch1,Bloch2} has taken
place. Greiner et al.  \cite{Bloch1}, to name one important example, succeeded
in experimentally driving a quantum phase transition between a superfluid and
a Mott-insulating phase in bosonic systems. This experimental progress has
revived the interest in the Bose-Hubbard model [Eq. (\ref{bh})] as a generic
Hamiltonian for strongly correlated bosons, by which the quantum phase
transition can be described \cite{Jaksch}. The Bose-Hubbard Hamiltonian has
been used previously in condensed matter physics to study the adsorption of
noble gases in nanotubes \cite{Chen}, or Cooper pairs in superconducting films
with strong charging effects \cite{Cha91,Fischer90}.  In this context a lot of
work has already been done to characterize the quantum phase transition, the
statistics, and the low-energy excitations of the Bose-Hubbbard model
\cite{Sachdev,FisherFisher1989}. However, new interesting questions arise now
due to the good tunability of the experiments with optical lattices.  In
particular, it becomes possible to study time-dependent processes such as
driven quantum phase transitions \cite{Bloch1}. A theoretical understanding of
such phenomena is challenging, since the characteristics of the superfluid
phase --- where the atoms tend to delocalize throughout the lattice and large
fluctuations in the local density exist ---, and the Mott-insulating phase ---
where the number fluctuations decrease, and a gap in the excitation spectrum
opens -- must be covered at the same time.  Both regions are separated by a
non-analyticity of the spectrum, which implies that a perturbative study
\cite{freericks94,freericks96} works best in strong coupling limit, while a
Hartree-Fock-Bogoliubov mean field works best in the superfluid regime. In
addition it is possible to develop a mean field theory \cite{Stoof,Rey} based
on a Gutzwiller ansatz \cite{Jaksch}: this reproduces the mean field theory in
the superfluid limit, as well as the limit of infinite interaction, which
raises the hope that the theory also interpolates properly between these
limits.

In this paper we develop a variational description of the ground state of an
ensemble of cold atoms in an one-dimensional optical lattice. Our trial
wavefunction treats the connections between neighboring sites as entities
which decouple in the limit of infinitely large lattices. We apply this
technique first to the quantum rotor model ---which describes the lattice for
large and commensurate occupation per site \cite{Cha91}, and has also been
used to describe an array of Josephson junctions
\cite{FaziovonderZant2001}---, and next to the Bose-Hubbard Hamiltonian. The
accuracy of the variational theory in both models is confirmed by several
comparisons. In the quantum rotor case, we use a spin wave approximation in
the limit of weak interaction, and a first order perturbation theory in the
limit of weak tunneling. For the Bose-Hubbard model we compare against results
obtained applying the quasi-exact, numerical DMRG method to one-dimensional
lattices with up to 128 sites, and also with calculations based on the
Gutzwiller ansatz. Our conclusion is that the variational picture of
self-regulated connections between sites provides a rather cheap and simple,
but very good description of the local properties of the system in the
superfluid and insulator regimes, and a fairly good interpolation across the
quantum phase transition. It cannot describe, however, the algebraic decay
with distance of the off-diagonal elements of the one-particle density matrix.
The simplicity of the method suggests a possible generalization to higher
dimensionalities and other physical models.

The outline of the paper is as follows: In Sec. {\ref{sec-rotor}} we introduce
the quantum rotor model as a possible limit of the Bose-Hubbard Hamiltonian.
Next, information about the ground state of the quantum rotor model is
obtained variationally as the solution of a Mathieu equation. We can estimate
energies, correlation functions and length, and the variance of the density as
a function of the only free parameter. A comparison with perturbative
estimates demonstrates the accuracy of the method when computing local
properties. Since the quantum rotor model is only an approximate description
of the optical lattice, in Sec. {\ref{sec-coherent}} we develop a similar
variational theory for the Bose-Hubbard Hamiltonian. After bringing the
Bose-Hubbard Hamiltonian to an appropriate form, we can estimate the local
properties of its ground state. The variational solutions are compared in Sec.
{\ref{sec-dmrg}} with the results of DMRG studies of the Bose-Hubbard model. We
confirm that the variational method describes very well the local properties
of both the Mott-insulator and the superfluid regime, and provides a fairly
good interpolation across the phase transition.  Finally, in Sec.
{\ref{sec-end}} we summarize our results and comment on possible extensions.

% ============================================================

\section{ Quantum phase model}
\label{sec-rotor}

\subsection{ Relation to the Bose-Hubbard model}
\label{sec-basics}

In this section we show the equivalence of the Bose-Hubbard model
\begin{equation}
  H_{BH} = \sum_{j=1}^{M}\left[-J (a^\dagger_{j+1}a_j+a^\dagger_ja_{j+1}) +
    \frac{U}{2} a^\dagger_ja^\dagger_ja_ja_j\right]
  - \frac{U}{2} M \bar{n}(\bar{n}-1),\label{bh}
\end{equation}
and the quantum rotor model for large and integer occupation $\bar{n}$
\cite{Cha91}.  In our notation, $M$ is the number of lattice sites and
$N=\bar{n}M$ the number of atoms. Both the Bose-Hubbard model and the quantum
rotor model show a phase transition due to the interplay between the kinetic
term proportional to $J$ and the interaction term proportional to $U$. For
convenience we have subtracted the ground state energy in the perfect
insulator limit $U/J\to\infty$.

If we expand a configuration of the lattice using Fock states
\begin{equation}
  \ket{\psi} =
  \sum_{\vec n} c_{\vec n}\ket{\vec n} =
  \sum_{\vec n} c_{\vec n}\ket{n_1}\otimes\cdots\otimes\ket{n_M},
\end{equation}
and the number of particles per lattice site is large, $n_k > 1$, we may
approximate the hopping terms as follows
\begin{equation}
  \ad_la_j\ket{\psi} =
  \sqrt{\bar{n}(\bar{n}+1)} {\cal P} A^+_l A^-_j\ket{\vec{n}}
  +\ket{\Delta_{lj}}
  \label{aprox-kin}
\end{equation}
Here, $A^{\pm}\ket{n}=\ket{n\pm 1}$ are ladder operators and ${\cal P}$
projects on states with non-negative occupation numbers, $n_k\geq 0$. To lowest
order the error $\ket{\Delta_{lj}}$ is
\begin{equation}
  \ket{\Delta_{lj}} = \sum_{\vec n}c_{\vec n}
  \frac{(\bar n +1)(n_j-\bar{n}) + \bar{n}(n_l-\bar{n})}
  {2\sqrt{\bar{n}(\bar{n}+1)}}
  \ket{\vec n},
\end{equation}
and its norm is bound by
\begin{equation}
  \Vert\Delta_{lj}\Vert \leq
  \sqrt{\frac{\bar{n}^2+(\bar{n}+1)^2}{2\bar{n}(\bar{n}+1)}}
  \sigma_l,
\end{equation}
where $\sigma_l = \braket{(n_l - \bar{n})^2}$ is the variance in the number of
particles per lattice site. For the approximation (\ref{aprox-kin}) to be
valid, the uncertainty in the number of atoms must be small compared to the
mean value, $\bar{n} \gg \sigma$, and the interaction energy must exceed the
neglected terms, $U\bar{n}(\bar{n}-1) \gg J\sigma$.

Following the previous procedure the Bose-Hubbard model becomes
\begin{equation}
  H_{{QR}} =
  {\cal P}\sum_j \left[-2J\rho(A^+_{j+1}A^-_j + A^+_jA^-_{j+1})+
    \frac{U}{2}(A^z_j)^2\right]
  \label{H-qr}
\end{equation}
Here $\rho=\sqrt{\bar{n}(\bar{n}+1)}$ is approximately the density,
$A^z_j=a^\dagger_ja_j-\bar{n}$ is essentially the number operator and we have
used that $\sum_jA^z_j\ket{\psi}=0$ when we work with states that have a fixed,
commensurate number of particles. In the following we will define the energy
per lattice site as
\begin{equation}
  \varepsilon \equiv \frac{1}{M}\langle{H_{{QR}}}\rangle.
  \label{eq:mu}
\end{equation}

Since the physically interesting states will be concentrated around large
occupations, $n_k=\bar{n}$, the usual step now is to drop the projector,
${\cal P}$, and move to the basis of phase states, defined by
\begin{equation}
  \braket{\vec{n}|\vec{\phi}}=
  e^{i\vec{n}\cdot\vec{\phi}}(2\pi)^{-M/2},\quad
  \vec\phi\in[-\pi,\pi]^{\otimes M}.
  \label{def-1}
\end{equation}
In doing so, we obtain the identification $A^\pm_j \to e^{\pm i\phi_k}$ and
$A^z_j \to -i\partial/\partial\phi_j$, which produces the usual representation
of the quantum rotor model
\begin{equation}
  H_{{QR}} =
  \sum_j \left[-2J\rho\cos(\phi_j-\phi_{j+1})-
    \frac{U}{2}\frac{\partial^2}{\partial\phi_j^2}\right],
  \label{H-qr-2}
\end{equation}
with the associated state writen as
\begin{equation}
  \ket{\psi} =
  (2\pi)^{-M/2} \int d^M\phi e^{i\bar{n}\sum\phi_k}\Psi(\vec\phi)
  \ket{\vec \phi}.\label{phase}
\end{equation}
A similar derivation is possible using path integrals \cite{path-integral}.
%% In the following we will define the energy per lattice site, excluding the
%% constant of Eq. (\ref{cosines}), as
%% \begin{equation}
%%   \varepsilon \equiv \frac{1}{M}\langle{H_{{QR}}}\rangle -
%%   \frac{U}{2}\bar{n}(\bar{n}-1).
%%   \label{eq:mu}
%% \end{equation}

\subsection{ Variational ansatz}

In this section we estimate the properties of the ground state of
$H_{{QR}}$ variationally. Due to the previous splitting (\ref{phase}),
any wavefunction $\Psi(\vec{\phi})$ can only depend on the phase difference
between neighboring wells, $\xi_j=\phi_{j+1}-\phi_j$. These new quantum
variables describe the connections between neigboring sites. In the limit of
large lattices it seems reasonable to assume that these connections become
independent from each other adopting the product state
\begin{equation}
  \Psi(\vec{\phi}) = \Pi_{j=1}^M h(\phi_j-\phi_{j+1}).\label{ansatz-phase}
\end{equation}
This representation becomes exact in the Mott-insulating limit, $U/J\to
\infty$, where $h_{mott}(\xi)=1$, and in the superfluid limit, $U/J\to 0$,
where $h_{sf}(\xi) = \sum_{n\in\mathbb{Z}}\delta(\xi-2\pi n)$, as can be
verified by direct substitution in Eq. (\ref{phase}).

\begin{figure}
  \centering
    \resizebox{0.5\linewidth}{!}{\includegraphics{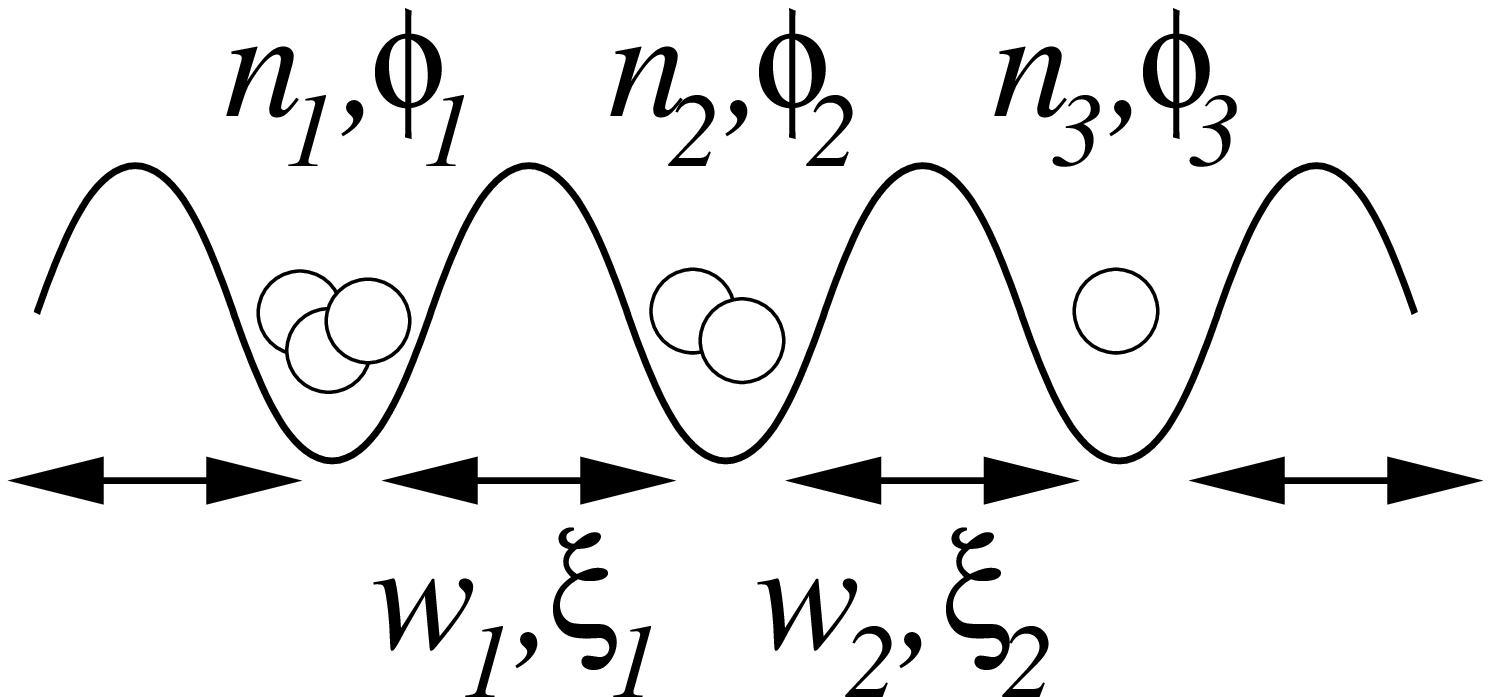}}
  \caption{
    \label{fig-m-numbers}
    Instead of working directly with the population of each well, $n_k$, we can
    use other quantum numbers, $w_k$, defined by the relation $n_k = w_k -
    w_{k-1} + \bar{n}$, and which behave like a set of chemical potentials
    acting on the barriers that connect neigboring sites.}
\end{figure}

Even though the phase representation is the best one to find a trial
wavefunction, it is not the optimal one for performing computations. It is
instead more convenient to work with the variables which are conjugate to the
phase differences $\xi_j=\phi_{j+1}-\phi_j$. These are the new quantum
numbers, $w_k$, given by
\begin{equation}
  n_k = w_k - w_{k-1} + \bar{n}.\label{quantum-numbers}
\end{equation}
In terms of these numbers, the ansatz (\ref{ansatz-phase}) reveals itself as a
simple product wavefunction
\begin{equation}
  \ket{\psi} = \ket{\tilde{h}}^{\otimes(M-1)}
  =\sum_{\vec{w}}\tilde{h}_{w_1}\cdots
  \tilde{h}_{w_{M-1}}\ket{w_1}\otimes\cdots\otimes\ket{w_{M-1}},\label{ansatz}
\end{equation}
with coefficients given by the Fourier transform
\begin{equation}
  \tilde{h}_m = \int h(\xi)e^{im\xi}d\xi.\label{equivalence}
\end{equation}
As sketched in Fig. \ref{fig-m-numbers}, the $w_k$ play the roles of chemical
potentials which are established between different wells: the difference
between the potentials on the extremes of a site gives the fluctuations over
the mean and commesurate occupation $\bar{n}$. In this picture
\begin{equation}
  H_{{QR}} = \sum_{k=1}^{M-1}\left[-2J\rho (\Sigma^+_k + \Sigma^-_k)
    + \frac{U}{2}(\Sigma^z_k-\Sigma^z_{k-1})^2\right],
  \label{H-qr-3}
\end{equation}
where $\Sigma^{\pm}\ket{w}=\ket{w\pm 1}$ are new infinite-dimensional ladder
operators and $\Sigma^{z}\ket{w}=w\ket{w}$.

By minimizing the energy associated with $H_{{QR}}$ over all states
within a given ansatz we can both obtain an upper bound to the energy of the
ground state and approximate its wave function. A simple computation with our
product ansatz leads to the result [compare (\ref{eq:mu})]
\begin{equation}
  \varepsilon[\tilde{h}]
  \simeq -4J\rho \mathrm{Re}\braket{\Sigma^+}
  + U\braket{(\Sigma^z)^2}
  - U \braket{\Sigma^z}^2,
\end{equation}
where the expected values are computed over a single connection,
$\braket{\Sigma^z} = \sum_w w |h_w|^2$, and the wavefunctions are assumed to be
normalized, $\sum_w|h_w|^2=1$. Since the stationary states have a well defined
parity, $\tilde{h}_{(-w)} = (-1)^P \tilde{h}_w$, the optimal variational state
must satisfy the linear equation
\begin{equation}
  -2J\rho(\tilde{h}_{j+1}+\tilde{h}_{j-1})+ U j^2 \tilde{h}_j =
  \varepsilon_{{est}} \tilde{h}_j,\label{mathieu-discrete}
\end{equation}
which is nothing but the Fourier transform of a Mathieu equation
\begin{equation}
  \left[-\frac{U}{2}\frac{\partial^2}{\partial\xi^2}-2J\rho\cos(\xi)\right]
  h(\xi) = \varepsilon_{{est}} h(\xi),\label{mathieu}
\end{equation}
The estimated ground state energy per site is given by the lowest eigenvalue of
either equation.

Using the product ansatz and the same approximations required to derive
$H_{{QR}}$, we can also compute other properties of the ground state.
For instance, the variance of the number of atoms per lattice site
\begin{eqnarray}
  \sigma_j^2 &=& \braket{(\ad_ja_j - \bar{n})^2} =
  2\braket{(\Sigma^z)^2}.\label{squeezing}
\end{eqnarray}
and the correlation functions
\begin{eqnarray}
  \braket{\ad_{j+1}a_j} &=& \rho \braket{\Sigma^-_j} \equiv \rho \gamma_1,
  \label{corr-mathieu}\\
  \braket{\ad_{j+l}a_j} &=&
  \rho \left\langle\prod_{k=j}^{j+l}\Sigma^-_k\right\rangle =
  \rho \gamma_1^l,
  \label{decay}
\end{eqnarray}
which decay exponentially with the distance. This implies that the
ansatz (\ref{ansatz}) only describes properly the decay of the correlations in
the Mott-insulating regime, since the correlations in the superfluid regime
follow a power law decay. However, as we will see below, local magnitudes
($\sigma,\, \gamma_1,\,\varepsilon\ldots$) are properly estimated even if
long--range ones are not.

\begin{figure}
\centering \resizebox{\linewidth}{!}{\includegraphics{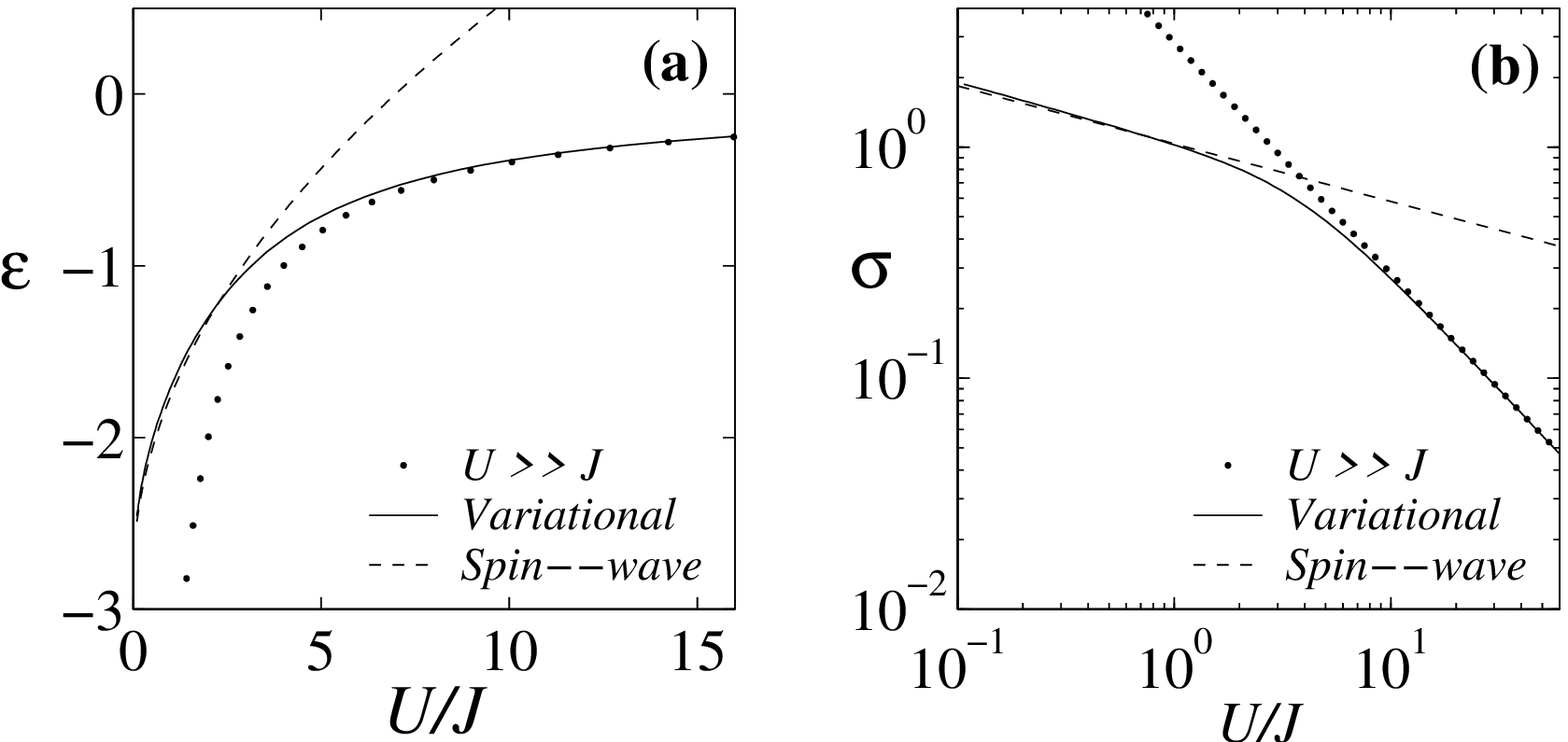}}
\caption{\label{fig-qr}
  Estimates for (a) energy energy per lattice site and (b) density
  fluctuations of the quantum rotor Hamiltonian (\ref{H-qr}) obtained with
  the variational method (solid), and perturbative calculations for $U\ll{}J$
  (dashed) and $U \gg J$ (dots).}
\end{figure}

We have solved Eq. (\ref{mathieu-discrete}) numerically in a truncated space.
The results are summarized in Fig. \ref{fig-qr}, where we also plot reference
estimates arising from two other analytical methods. In the limit $U \gg J$ we
compare with a first order perturbative calculation around the solution $\ket{\psi}=1$,
which is possible thanks to the energy gap of order ${\cal O}(U)$ in the
excitation spectrum. In the limit $J \gg U$ we rather use a spin wave or
harmonic approximation in which the cosine-term of the Hamiltonian
$H_{{QR}}$ is expanded up to second order in the phase difference
between neighboring sites [see Sec. {\ref{appendix-c}}].  This approximation is
valid in the superfluid regime, where the phase does not vary much between
neighboring wells.  From the graphical comparison we see that, as expected, the
variational wavefunction provides a fairly accurate description of the ground
state of the quantum rotor model in both the superfluid and insulating limits.
As a side note, we must remark that this ground state has a divergent
fluctuation of the number of particles per site as $J \to 0$.  This disagrees
from the expected behavior of the ground state of the original Bose-Hubard
Hamiltonian, and it reminds us that $H_{{QR}}$ can only model the atomic
ensemble when the variance, $\sigma$, is small compared to the mean occupation
number, $\bar{n}$.

\subsection{ Harmonic approximations to the quantum phase model}
\label{appendix-c}

In the limit $J\gg U$ it is possible to estimate the ground state of the rotor
model (\ref{H-qr}) analytically. Since we are deep in the superfluid regime,
the wavefunction will be concentrated around the line
$\phi_1=\phi_2=\ldots=\phi_M$, and we can approximate
\begin{equation}
  H_{{QR}}
  \simeq \sum_j \left[-\frac{U}{2}\frac{\partial^2}{\partial\phi_j^2}
    - J\rho(\phi_j-\phi_{j+1})^2\right].
\end{equation}
If we remove the periodic boundary conditions on $\phi_i$ and change variables,
the preceding Hamiltonian may be diagonalized, $H = \sum_k \omega_k
\left(A^\dagger_kA_k + \frac{1}{2}\right)$, with frequencies given by
$\omega_k = \sqrt{8J\rho U} \; |\sin(\pi k/M)|$, where $k$ is an integer in the
range $-M+1<2k<M$ which labels the different values of the momentum in the
lattice. The ground state energy [Fig. \ref{fig-gutz}(a)] may be estimated as
the zero-point energy of the harmonic oscillator. For large $M$, the sum over
$k$ may be replaced with an integral, giving
\begin{equation}
  E_g \simeq \frac{2M}{\pi}\sqrt{2J\rho U}.
\end{equation}
The variance of the number of particles is related to the expectation value
of the momentum using the same procedure as above
\begin{equation}
  \sigma \simeq \frac{1}{\pi}\sqrt{\frac{8 J \rho}{U}}.
\end{equation}
%% The correlations are a bit more complicated to estimate. We have to express
%% the expectation value $c_l = \braket{\exp(i\phi_{j+l}-i\phi_j)}$ in terms of
%% the position operator $A_k + A_k^\dagger$. More precisely,
%% \begin{eqnarray}
%%   c_l &=& \left\langle\exp
%%     \left[i\sum_k\alpha_k\tfrac{1}{\sqrt{2}}(A_k+A_k^\dagger)\right]\right\rangle\\
%%   \alpha_{k\geq0} &=& -\frac{2\sqrt{2}}{\sqrt{M}}
%%   \sin^2\left(ql\right)
%%   \left(\frac{U}{8J\rho\sin^2(q)}\right)^{1/4},\nonumber\\
%%   \alpha_{k<0} &=& \sqrt{\frac{2}{M}}
%%   \sin\left(ql\right)
%%   \left(\frac{U}{8J\rho\sin^2(q)}\right)^{1/4},\nonumber
%% \end{eqnarray}
%% where $q=\pi |k| /M$ ranges approximately from $0$ to $\pi/2$. These sums
%% may be computed in the limit of large lattices for certain values of $l$.
%% For instance, if $l=1$, we replace the sum over $k$ with an integral over
%% $q$ and get
%% \begin{equation}
%%   c_1 = e^{-\sum_k|\alpha_k|^2/4} =
%%   \exp\left(\frac{2}{3\pi}\sqrt{\frac{2U}{J\rho}}\right).
%% \end{equation}

% ======================================================================

\section{ The Bose-Hubbard model}
\label{sec-coherent}

In this section we apply to the Bose-Hubbard Hamiltonian the techniques that
were developed in Sec. {\ref{sec-rotor}}. We will do it in three
steps: First we will develop a phase representation which is valid for all
occupation numbers.  Next we will prove that this representation is equivalent
to a similarity transformation of the Hamiltonian which brings it to a form
similar to (\ref{H-qr}), at the price of losing Hermiticity. Finally we will
show how to implement the ansatz of independent connections (\ref{ansatz}) to
produce estimates for the usual set of observables ($\varepsilon,\,\gamma_1,\,\sigma$),
which are to be validated with DMRG calculations.

\subsection{ Coherent states}

The phase coherent states $\ket{\phi}$ are defined by
\begin{equation}
  \label{coh-states}
  \braket{n|\phi} = e^{in\phi}/\sqrt{n!}.
\end{equation}
Unlike the phase states defined in Sec. {\ref{sec-rotor}}, they are not
orthogonal to each other, $\braket{\phi|\theta} =
\exp\left[e^{i(\theta-\phi)}\right]$, but they form a complete basis, so that
an expansion like (\ref{phase}) is still possible. A nice property of the
coherent states is that we can rewrite the operators $a$, $a^\dagger$,
$a^\dagger a$, etc, in terms of the phases very easily. For instance,
$a\ket{\phi} = e^{i\phi}\ket{\phi}$, $a^\dagger\ket{\phi} =
-ie^{-i\phi}\frac{\partial}{\partial\phi}\ket{\phi}$, and $a^\dagger
a\ket{\phi} = -i\frac{\partial}{\partial\phi}\ket{\phi}$.  Using this
representation, we obtain an effective Hamiltonian for the wavefunction
$\Psi(\vec{\phi})$, i.e. $H_{{BH}}\ket{\psi} = (2\pi)^{-M/2} \int
d^M\phi e^{i\bar{n}\sum\phi_k}[H_{{coh}}^t\Psi(\vec\phi)] \ket{\vec
  \phi}$, which is of the form
\begin{equation}
  H_{{coh}}^t
  = -J\sum_{\langle i,j\rangle}\left[2(\bar{n}+1)\cos(\phi_i-\phi_j)
  -ie^{i(\phi_i-\phi_j)}\frac{\partial}{\partial\phi_j}\right]
  + \frac{U}{2}\sum_j \left(-\frac{\partial^2}{\partial\phi_j^2}\right).
  \label{H-phase-2}
\end{equation}
Here $H_{{coh}}^t$ stands for the tranpose of $H_{{coh}}$. This operator was
already used in Ref. \cite{Anglin01} to study the Bose-Hubbard model with only
two sites. On the one hand, it is a non-Hermitian operator\footnote{The
  hermiticity of $H_{BH}$ is maintained due to an implicit projection that
  takes place when we reconstruct the state $H_{BH}\ket{\psi}$ from
  $H_{coh}^t\Psi(\vec{\phi})$ (See Ref. \cite{Anglin01}).} and we cannot do a
simple variational study.  On the other hand the Hamiltonian still depends on
the phase differences, and it is reasonable to look for approximate
eigenstates which have the form (\ref{ansatz-phase}). This will be done in the
following section. However, since working with phase variables is
inconvenient, we will develop a representation similar to that of Eq.
(\ref{H-qr-3}) in the following.

\subsection{ Variational procedure for non-Hermitian operators}

In this section we will find the best variational function which has the
product form of Eq. (\ref{ansatz-phase}). However, as it happened in Sec.
{\ref{sec-rotor}}, instead of working with phase variables it is more convenient
develop a representation of the Bose-Hubbard Hamiltonian in terms of
connections. This is once more a two-steps process. First we use a similarity
transformation suggested by the definition of the coherent states
\begin{equation}
  O\ket{\vec{n}} = \prod_{k=1}^M \sqrt{n_k!}\ket{\vec{n}}.
\end{equation}
Since $Oa_jO^{-1} = A^-_j$ and $Oa^\dagger_jO^{-1}=A^z_jA^+_j$, we find
\begin{eqnarray}
  \label{H-phase}
  H_{{coh}} &=& OH_{{BH}}O^{-1}\\
  &=& -J\sum_{\braket{i,j}}(A^z_i+\bar{n})A^+_iA^-_j +
  \frac{U}{2}\sum_j (A^z_j)^2. \nonumber
\end{eqnarray}
The Hamiltonians (\ref{H-phase-2}) and (\ref{H-phase}) are equivalent: while
one is defined in terms of phase variables, the other one is defined with
occupations numbers, and both are related by a Fourier transform. The second
and final step is to rewrite everything in terms of connections, using the
quantum numbers from Eq. (\ref{quantum-numbers}), and the relations
$\Sigma^x=\Sigma^+ +\Sigma^-$, $\Sigma^y=i(\Sigma^--\Sigma^+)$. The result is
a decomposition of the Hamiltonian
\begin{eqnarray}
  \label{H-phase-3}
  H_{{coh}} &=& H_1+H_2,\\
  H_1 &=& \sum_j\left[-J \bar{n}\Sigma^x
  + iJ \Sigma^z_j\Sigma^y_j + U(\Sigma^z_j)^2\right],\nonumber\\
  H_2 &=& \sum_j\left[J(\Sigma^z_{j-1}\Sigma^+_j-\Sigma^z_{j+1}\Sigma^-_j)+
  U \Sigma^z_j\Sigma^z_{j+1}\right],\nonumber
\end{eqnarray}
into terms which are local, $H_1$, and terms which involve pairs of
connections, $H_2$.

For the quantum rotor model we proved that the optimal product wavefunction was
an eigenstate of a Hamiltonian which did not couple connections, like $H_1$.
The difference now is that, since the operator $H_{{coh}}$ is not
Hermitian, we cannot establish a variational principle and that proof is no
longer valid. Nevertheless, we will again propose a variational ansatz which is
an eigenstate of the local operator $H_1\ket{\tilde h}^{\otimes M} =
M\varepsilon_{{est}} \ket{\tilde h}^{\otimes N}$.  Using the
following equality
\begin{equation}
  \varepsilon_0 =
  \min_{\psi\neq 0}
  \frac{\braket{\psi|H_{{BH}}|\psi}}{\Vert\psi\Vert^2}
  = \min_{\chi\neq 0}
  \frac{\braket{\chi|O^{-2}H_{{coh}}|\chi}}{\braket{\chi|O^{-2}|\chi}},
\end{equation}
and the product ansatz $\ket{\chi}=\ket{\tilde h}^{\otimes M}$, we arrive to an
upper bound for the lowest eigenvalue of the Bose-Hubbard Hamiltonian,
expressed in terms of the non-Hermitian one
\begin{equation}
  \label{var-non-H}
  \varepsilon_0 \leq \varepsilon_{\mathrm{est}} +
  \frac{1}{N}\frac{\bra{\tilde h}^{\otimes M}O^{-2}H_2\ket{\tilde
  h}^{\otimes M}}{\bra{\tilde h}^{\otimes M}O^{-2}\ket{\tilde
  h}^{\otimes M}} \equiv \varepsilon_{{est}} + \Delta\varepsilon_{{est}}.
\end{equation}

\begin{figure}
  \centering
    \resizebox{\linewidth}{!}{\includegraphics{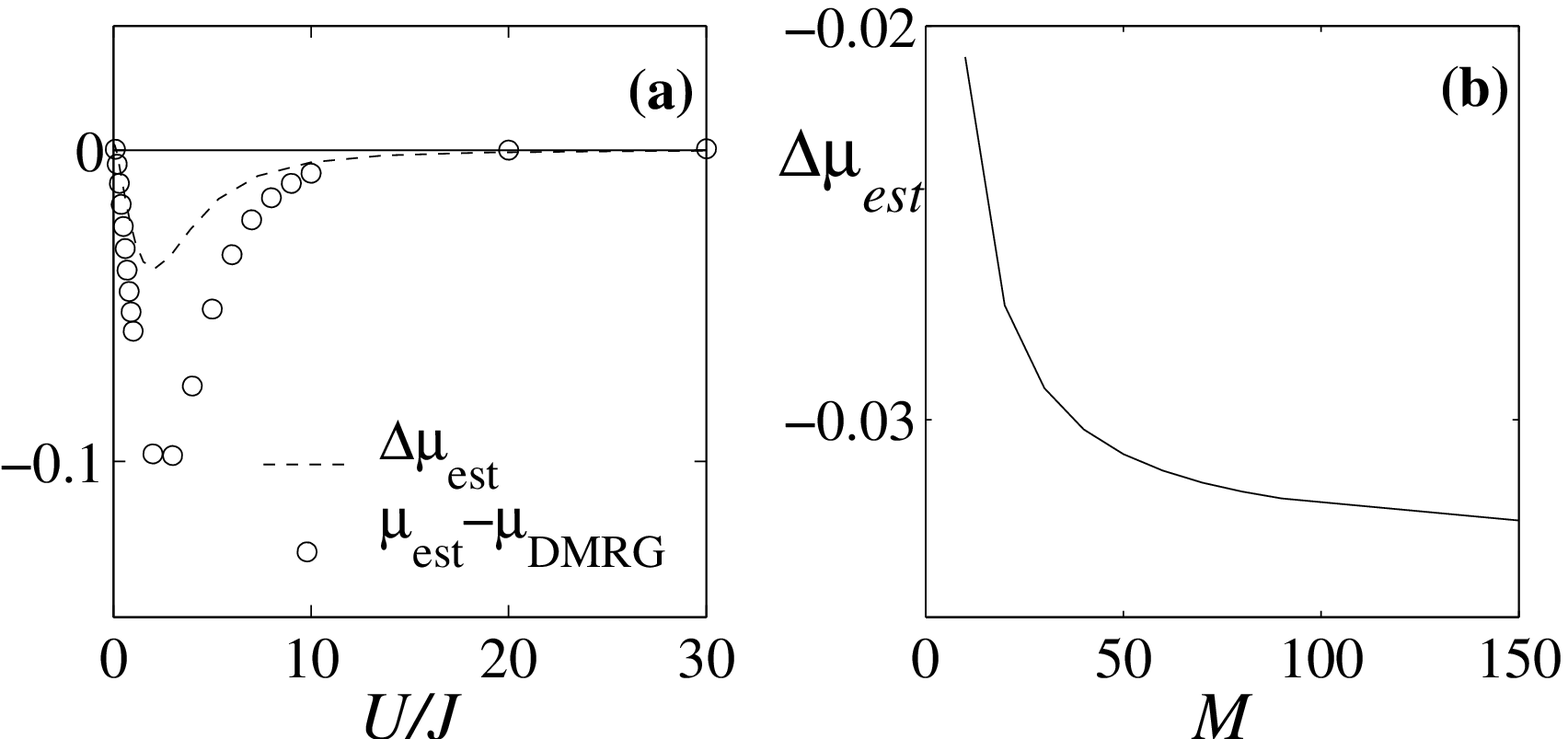}}
  \caption{
    \label{fig-correction}
    The energy of the product ansatz contains a contribution from each
    connection, $\varepsilon_{{est}}$, plus the interaction between neighbouring
    connections, $\Delta\varepsilon_{{est}}$. In Fig. (a) we show that
    $\Delta\varepsilon_{{est}}$ (dash) is actually negative, and improves the
    estimate $\varepsilon_{{est}}$ moving it towards the exact value,
    $\varepsilon_{{DMRG}}$ (circles). Everything has been computed for
    $\bar{n}=1$. In Fig. (b) we show that the correction
    $\Delta\varepsilon_{{est}}$ does not change much for large lattices.}
\end{figure}

The way to use this variational principle is as follows. First, for a given $J$
and $U$ we compute the lowest eigenstate of $H_1$ and this way obtain $\tilde
h$. After the equivalence (\ref{equivalence}), finding the ground state of the
local Hamiltonian $H_1$ becomes equivalent to solving a modified Mathieu
equation
\begin{equation}
  \label{anglin}
  \left[-U\frac{\partial^2}{\partial\xi^2}-2J(\bar{n}+1)\cos(\xi)
    -2J\sin(\xi)\frac{\partial}{\partial\xi}\right]h = \varepsilon_{{est}}h,
\end{equation}
which describes exactly the static properties of a pair of sites with open
boundary conditions \cite{Anglin01}. Once we have $\varepsilon_{{est}}$ we must
still compute the correction $\Delta\varepsilon_{{est}}$ using a rather
straightforward expansion which is shown in Sec. {\ref{sec-bounds}}.
Surprisingly, $\Delta\varepsilon_{{est}}$ happens to be negative, so that it is
actually an improvement over the simple estimate given by $\varepsilon_{{est}}$
[See Fig.  \ref{fig-correction}].

From the optimal variational state, $\ket{\psi} = O^{-1}\ket{\tilde h}^{\otimes
  M}$, and the estimate for the energy, $\varepsilon_{{var}} =
\varepsilon_{{est}} + \Delta\varepsilon_{{est}}$, we may compute other
observables. For the density fluctuations and nearest neighbor correlations we
use the virial theorem
\begin{eqnarray}
  \braket{a^\dagger_{j+1}a_j} &=& \frac{\partial}{\partial
    J}\varepsilon_{{var}},
  \\
  \sigma^2 &=& \frac{\partial}{\partial U}\varepsilon_{{var}}- \bar{n}^2,
\end{eqnarray}
whereas for other properties one has to evaluate numerically the matrix
products shown in Sec. {\ref{sec-bounds}}. This allows us to prove that for the
product states $\ket{\tilde h}^{\otimes M}$ correlations decay exponentially,
opposite to what is expected in the superfluid phase, whose correlations should
decay algebraically. Nevertheless, as we will see next, this family of states
does estimate accurately the local properties of the optical lattice.

% ======================================================================

\begin{figure}[t]
  \centering
    \resizebox{\linewidth}{!}{\includegraphics{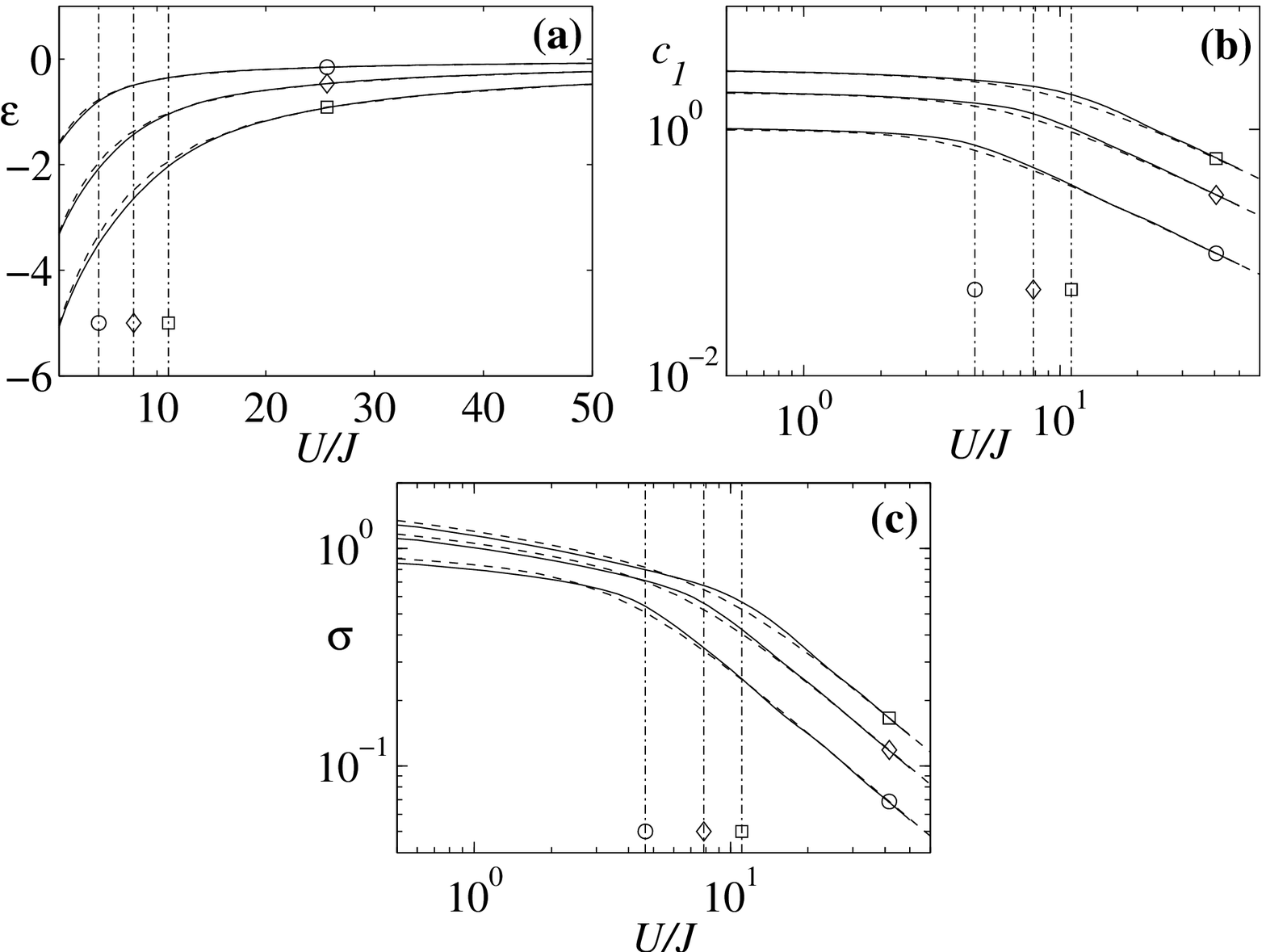}}
  \caption{
    \label{fig-dmrg}
    (a) The ground state energy per site, $\varepsilon$, (b) nearest neighbor
    correlation, $c_1=\braket{\ad_{j+1}a_j}$, and (c) variance of the number of
    atoms per site, $\sigma^2=\braket{(n_j - \bar{n})^2}$.  Plots (b) and (c)
    use a log-log scale. The results of the DMRG (solid line) are obtained on a
    system with 128 sites, a maximum occupation number of 9 bosons per site and
    a reduced space of states of about $200$ states.  The estimates from the
    variational theory are plotted using dashed lines.  The vertical lines mark
    the location of the phase transition according to \cite{freericks96}. The
    mean occupation numbers are denoted with circles ($\bar{n}=1$), diamonds
    ($\bar{n}=2$) and boxes ($\bar{n}=3$).}
\end{figure}

\subsection{ Comparison to DMRG results}
\label{sec-dmrg}

We will now compare the results for the ground state energy, the correlation
functions, and the variance of the particle number provided by the two
variational ansatz, (\ref{mathieu}) and (\ref{anglin}), and the Gutzwiller
ansatz \cite{vanOtterloSchoen1995} with those obtained by DMRG studies of the
Bose-Hubbard model. The DMRG, developed 1992 by White \cite{white92, white93}
in the area of condensed matter theory, is a very powerful numerical tool to
investigate static and dynamic properties of strongly correlated
quasi-one-dimensional spin, fermionic or bosonic quantum systems.  The DMRG is
an essentially quasi-exact numerical method. The fundamental ideas stem from
real space renormalization methods: the system size is grown iteratively while
the (exponentially diverging) size of the Hilbert space is kept constant by
decimation. Hereby one tries to retain only that subset of states that is
essential to describe the physical quantity under consideration. In DMRG these
are expectation values with respect to low-lying states (``target states''),
and in particular with respect to the ground state wave function.

DMRG builds up the system linearly: at each growth step, suitable density
matrices for the target states are derived that yield information on the
relevance of Hilbert space states. Building on this information, the states
and operators are projected onto Hilbert subspaces of fixed dimension $M$
containing the most relevant states. $M$ is chosen to be small enough to be
handled numerically, but large enough to obtain the desired accuracy;
numerical results can be extrapolated in $M$ to the exact limit of infinite
$M$ in the thermodynamic limit. However, results presented here have converged
for the largest $M$ considered and no further extrapolation was necessary.

Details on the DMRG method may for example be found in \cite{bookDMRG}. In the
case of the Bose-Hubbard model the DMRG has been used to study properties of
the system \cite{KuehnerMonien2000,diplomKuehner,RapschZwerger1999}. The
results of DMRG agree very well with exact diagonalization results for small
systems, with quantum Monte-Carlo simulations e.g.
\cite{BatrouniZimanyi1995,BatrouniScalettar1992,NiyazBatrouni1994,
  ProkofevTupitsyn1998}, and with 13th order perturbation theory
\cite{ElstnerMonien1999}.

We have used the DMRG to study the properties of the ground state of the
Bose-Hubbard model on one-dimensional lattices with 128 sites, and
commensurate fillings $\bar{n}=1,2,$ and $3$. In Figs. \ref{fig-dmrg}(a-c) we
show the results for the mean energy per site $\varepsilon$, the nearest
neighbor correlations, $c_1=\braket{\ad_{j+1}a_j}$, and the variance $\sigma$
of the density, calculated both with the DMRG and with the variational
estimates developed above. As expected, there are no indications of the phase
transitions in these quantities, neither in the variational results nor in the
numerical solutions.  Rather, an inflexion of the nearest neighbor correlation
points out the location of the superfluid-insulator transition which lies
roughly between $3\bar{n}$ and $4\bar{n}$ (see \cite{KuehnerMonien2000} and
ref. therein). The agreement of the two methods is fairly good above the phase
transition and below it.

\begin{figure}[t]
  \centering
    \resizebox{\linewidth}{!}{\includegraphics{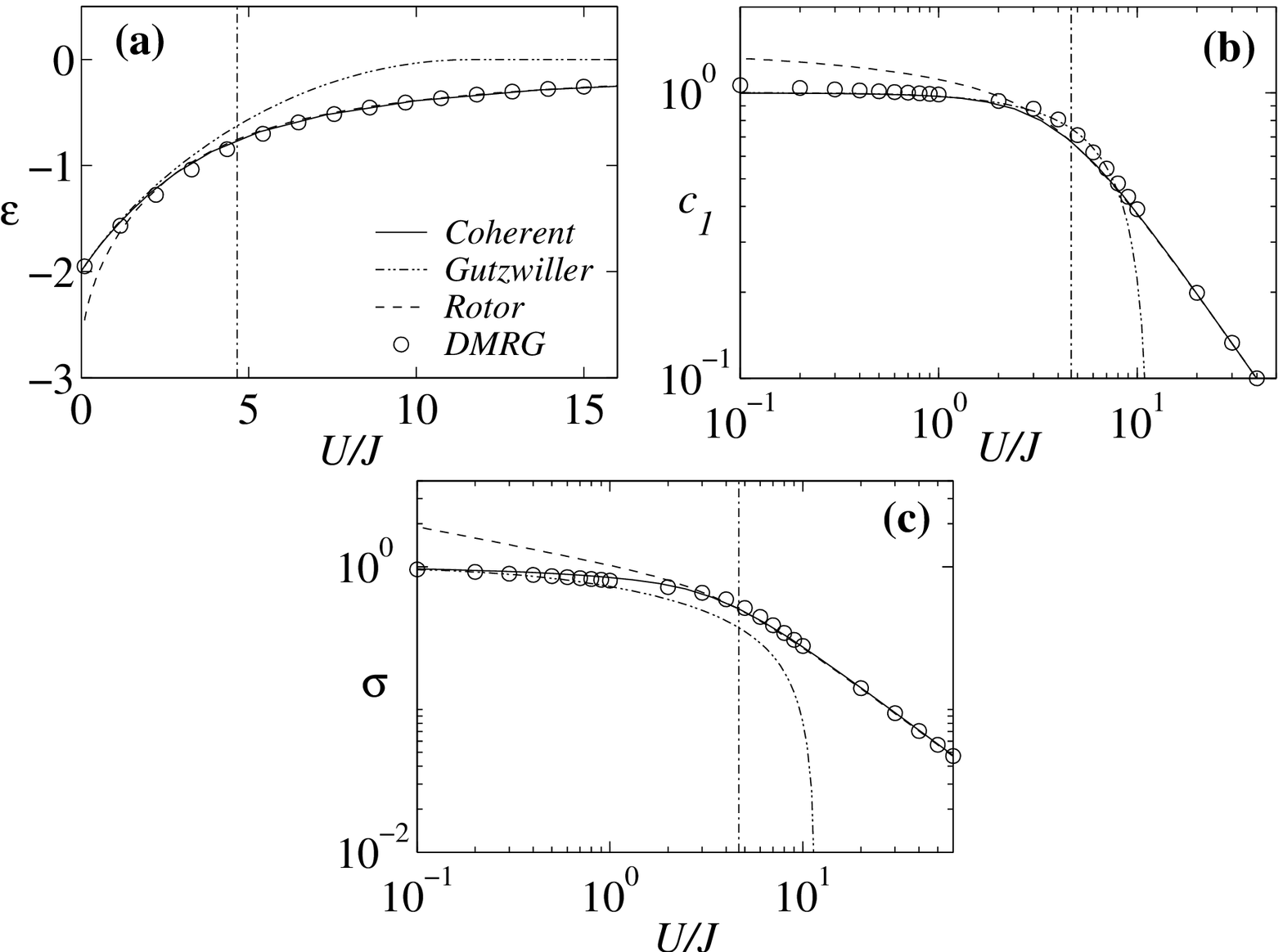}}
  \caption{
    \label{fig-gutz}
    (a) The ground state energy per site, $\varepsilon$, (b) nearest neighbor
    correlation, $c_1=\braket{\ad_{j+1}a_j}$, and (c) variance of the number of
    atoms per site, $\sigma^2=\braket{(n_j - \bar{n})^2}$. Plot (b) and (c) are in
    log-log scale. Using filling factor $\bar{n}=1$, we show results from the
    variational model for the Bose-Hubbard model using phase coherent states (solid), the
    quantum rotor model (dashed), the Gutzwiller ansatz for the Bose-Hubbard
    Hamiltonian (dots) and DMRG (circles). Vertical dash-dot lines mark the
    location of the phase transition according to \cite{freericks96}.}
\end{figure}

A more detailed comparison is provided in Fig. \ref{fig-gutz} for the case
$\bar{n}=1$. In this figure we plot together results from the DMRG, the
variational ansatz derived above, the quantum rotor model and the well-known
Gutzwiller ansatz. The Gutzwiller ansatz \cite{RokhsarKotliar1991,Jaksch} is a
variational ansatz which reduces the wave function to a product of single-site
wave functions, $\left|\Psi_G\right\rangle = \Pi_{j=1}^M \left| \Phi_j
\right\rangle$, where $\left| \Phi_j\right \rangle = \sum_{m=0}^{\infty}
f_m^{(j)} \left| m_j \right\rangle$ and $f_m^{(j)} $ are constants. Such a
wavefunction cannot be used in the one-dimensional Mott insulator regime,
because a perturbative study of the Gutzwiller ansatz shows that the
corrections of order ${\cal O}(J/U)$ are lost and all correlations become zero.
However this ansatz gives good results in the superfluid regime, where the
long-range order is well described by $\ket{\Psi_G}$, and we can use these
results and those of the DMRG to assert the accuracy of our variational
estimates. As Fig. \ref{fig-gutz} shows, in the Mott insulator regime, the DMRG
results agrees perfectly with our variational theory for the Bose-Hubbard model
and for the quantum rotor model. Close to the phase transition is the point at
which the quantum rotor model no longer describes well the atoms in the optical
lattice due to the growth of density fluctuations. At this point we also
observe a small disagreement between the DMRG and the coherent states, which is
due to the growth of long range correlations and vanishes as we get deeper into
the superfluid regime.

\subsection{ Numerical evaluation of the upper bound}
\label{sec-bounds}

In this section we will show how to compute the corrections to the local
energy, $\Delta\varepsilon_{{est}}$, from Eq. (\ref{var-non-H}). We basically
need a method to compute expectation values of the operator $O^{-2}$ around
product states which have the form
\begin{equation}
  \ket{\phi} = \ket{\tilde h}_1\cdots\ket{\tilde h}_{k-1}
  \ket{A\tilde h}_k\ket{B\tilde h}_{k+1}
  \ket{\tilde h}_{k+2}\cdots\ket{\tilde h}_M,
\end{equation}
in which at most two contiguous vectors are affected by two single-connection
operators.  For instance, this is the case of $H_2\ket{\psi}$, where $A$ and
$B$ are $\Sigma^+$, $\Sigma^-$ or $\Sigma^z$, and also of $\ket{\psi}$, where
$A$ and $B$ are just the identity. After some manipulations it is possible to
write, for the optimal variational state $\ket{\psi}=O^{-1}\ket{\tilde
  h}^{\otimes M}$,
\begin{eqnarray*}
  \braket{O^{-2}H_2}_\psi &=& \sum_{k=1}^{M-2}
  \vec{u}^{~t}O(HO)^{k-1}Z_kO(HO)^{M-k-1}\vec{u},\\
  Z_k &=& J(H_zOH_+-H_-OH_z)-U(H_zOH_z),\nonumber
\end{eqnarray*}
with the real matrices and vectors
\begin{eqnarray*}
  H_{ij} &=& |\tilde{h}_i|^2\delta_{ij},\\
  (H_\alpha)_{ij} &=& |\tilde{h}_i(\Sigma^\alpha\tilde{h})_i|^2\delta_{ij},
  \\
  O_{ij} &=& \left\{\begin{array}{cc}
      [(i-j+\bar{n})!]^{-1/2},&i-j\geq-\bar{n}\\
      0,&i-j<-\bar{n}\end{array}\right.,\\
  u_i &=& \delta_{i0},\\
  i,j&\in&\mathbb{Z},\;\alpha\in\{+,-,z\}.
\end{eqnarray*}

We have used this technique to compute numerically the correction
$\Delta\varepsilon_{{est}}$ using different lattice sizes and found small or no
differences for more than 30 sites. Intuitively, this is because in the limit
of large powers the matrices $(HO)^\alpha$ become projectors on the eigenvector
with the largest eigenvalue.

Using the same type of expansion we may compute other correlations
\begin{equation}
  \braket{a^\dagger_{k+\Delta}a_k}_\psi =
  \frac {\vec{u}^t O(HO)^{k-1}(H_-O)^\Delta (HO)^{M-k+1}\vec{u}}
  {{\vec u}^t O(HO)^M{\vec u}}.
\end{equation}
For large values of $\Delta$ and large lattices, the numerator will decay
exponentially as $\gamma_1^\Delta$, where $\gamma_1$ is the largest eigenvalue
of the matrix $H_-O$.

% ============================================================

\section{ Conclusions}
\label{sec-end}

In this work we have studied analytically and numerically the properties of the
ground state of an ensemble of bosonic atoms in an 1D optical lattice. For the
study of the atomic ensemble we have used both the quantum rotor model and the
Bose-Hubbard model. Exploiting the fact that in these models there exists only
nearest neighbor hopping and local interactions, we have developed a
variational wavefunction that may be used to easily estimate local properties,
such as the energy per well, the nearest neigbor correlations and the
fluctuations of the density. In the case of the quantum rotor model we have
verified our results with perturbative calculations around the strongly
interacting regime, and with a spin wave approximation around the superfluid
regime. In the case of the Bose-Hubbard model we have compared the variational
estimates with numerical results obtained using the DMRG technique for a maximum
 density of three atoms per well. We have concluded
that this procedure leads to fairly good estimates of local ground state
properties of both Hamiltonians, in both the superfluid and the insulator
regime, the largest disagreement being localized around the phase transition.
In future work we forsee the application of these methods to time dependent
problems.

\section*{Acknowledgments}

We thank W. Zwerger for fruitful discussions. J. J.  Garc\'{\i}a-Ripoll and J.
I. Cirac thank the Deutsche Forschungsgemeinschaft (SFB 631) and the
Kompetenznetzwerk Quanteninformationsverarbeitung der Bayerischen
Staatsregierung.  C. K. and U. S.  thank the Hess-Preis of the DFG and the
Studienstiftung des deutschen Volkes for financial support. Work at the
University of Innsbruck is supported by the Austrian Science Foundation, EU
Networks and the Institute for Quantum Information. P.Z. thanks the Max Planck
Institut f\"ur Quantenoptik for hospitality during his stay in Garching, and
thanks the Humboldt Foundation for support during this stay.

\end{document}